\documentclass[reprint,twocolumn,amsmath,amssymb,amsthm, dsfont,prl,nobibnotes]{revtex4-2}
\usepackage{times}
\usepackage[colorlinks=true,citecolor=blue,linkcolor=blue]{hyperref}
\usepackage{graphicx}
\usepackage{mathrsfs}
\usepackage{dcolumn}
\usepackage{bm}
\usepackage{hyperref}
\usepackage{tikz}
\usepackage{hhline}
\usepackage{float}
\usepackage{scalerel}
\usepackage{amsfonts}
\usepackage{bbm}
\usepackage{color}
\usepackage{tabularx}
\usepackage{multirow}

\usetikzlibrary{decorations.markings}

\makeatletter
\def\l@subsection#1#2{}
\def\l@subsubsection#1#2{}
\makeatother

\begin{document}
\tikzset{
    vertex/.style={fill,circle,draw,scale=0.3}
}

\newcommand{\tetrahedron}{
  \mathchoice
    {\includegraphics[height=1ex]{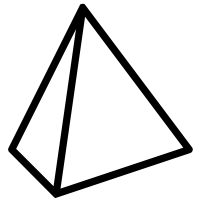}} 
    {\includegraphics[height=1ex]{tetrahedron}} 
    {\includegraphics[height=1.5ex]{tetrahedron}} 
    {\includegraphics[height=.5ex]{tetrahedron}} 
}


\def\bra#1{\mathinner{\langle{#1}|}}
\def\ket#1{\mathinner{|{#1}\rangle}}
\newcommand{\Ket}[1]{\vcenter{\hbox{$\displaystyle\stretchleftright{|}{#1}{\bigg\rangle}$}}}

\title{Symmetric Finite-Time Preparation of Cluster States via Quantum Pumps}

\author{Nathanan Tantivasadakarn}
\affiliation{Department of Physics, Harvard University, Cambridge, MA 02138, USA}
\author{Ashvin Vishwanath}
\affiliation{Department of Physics, Harvard University, Cambridge, MA 02138, USA}

\date{\today}

\begin{abstract}
It has recently been established that cluster-like states -- states that are in the same symmetry-protected topological phase as the cluster state -- provide a family of resource states that can be utilized for Measurement-Based Quantum Computation. In this work, we ask whether it is possible to prepare cluster-like states in finite time without breaking the symmetry protecting the resource state. Such a symmetry-preserving protocol would benefit from topological protection to errors in the preparation. We answer this question in the positive by providing a Hamiltonian in one higher dimension whose finite-time evolution is a unitary that acts trivially in the bulk, but pumps the desired cluster state to the boundary. Examples are given for both the 1D cluster state protected by a global symmetry, and various 2D cluster states protected by subsystem symmetries.  We show that even if unwanted symmetric perturbations are present in the driving Hamiltonian, projective measurements in the bulk along with feedforward correction is sufficient to recover a cluster-like state. 
\end{abstract}

\maketitle

\textit{Introduction.}
Symmetry-Protected Topological (SPT) states\cite{PollmannBergTurnerOshikawa12,ChenGuWen2011,ChenGuWen2011-2,Chenetal2013,ChenScience} are gapped states of matter that cannot be adiabatically connected to an unentangled product state without breaking the protecting symmetry. It has been recently realized that certain SPT states, and in some cases, entire SPT phases, can be leveraged to perform Measurement-Based Quantum Computation (MBQC)\cite{GottesmanChuang1999,BriegelRaussendorf2001,RaussendorfBriegel2001,RaussendorfBrowneBriegel2003}. The fact that these states cannot be smoothly connected to an unentangled product state without breaking a global symmetry in many cases implies an adequate entanglement structure of the state which is sufficient to perform MBQC.

So far, it is known that certain SPT phases host computational power throughout the entire phase: any state within that SPT phase can be used as a resource state\cite{Chenetal2009,WeiAffleckRaussendorf2011,Miyake2011,WeiAffleckRaussendorf2012,PrakashWei2015,MillerMiyake2015,NautrupWei2015,Raussendorfetal2017,Stephenetal2017,WeiHuang2017,ChenPrakashWei2018}. In one dimension, the canonical example is the cluster state, defined as the unique state which satisfies $Z_{i-1} X_i Z_{i+1}|\psi \rangle =+|\psi \rangle $ on a spin chain. The state enjoys a global $\mathbb Z_2^2$ symmetry and for any state within the same SPT phase, arbitrary quantum gates can be performed by choosing appropriate measurements, making the entire SPT phase universal\cite{Raussendorfetal2017,Stephenetal2017}. In higher dimensions, certain fixed points (with possibly finite regions around those fixed points) of SPT phases with global symmetries have been found to be universal \cite{NautrupWei2015,WeiHuang2017,MillerMiyake2016,ChenPrakashWei2018,MillerMiyake2018} for MBQC, but an SPT state with global symmetry whose entire phase is universal has yet to be found.

 On the other hand, there has been increased interest in subsystem symmetries due to their connections to fracton topological order in three spatial dimensions\cite{Chamon2005,Haah2011,Yoshida2013,VijayHaahFu2015,VijayHaahFu2016,Williamson2016,NandkishoreHermele19,PretkoChenYou20}. Unlike global symmetries, subsystem symmetries only act on a rigid sub-dimensional region, such as lines, planes or even fractals. It was recently realized that if one instead considers states protected by such symmetries, called subsystem SPT states\cite{Youetal2018,Devakuletal2018,DevakulWilliamsonYou2018,KubicaYoshida2018,DevakulShirleyWang2019,TantivasadakarnVijay2019,Tantivasadakarn20,Shirley2020}, then there are indeed examples where the entire phase can be used as a universal resource state\cite{Raussendorfetal2018,Stephenetal2018,DevakulWilliamson2018,DanielAlexanderMiyake20}. Serendipitously, these examples are again cluster states on various 2D lattices.
 
 \begin{figure}
    \centering
    \includegraphics[scale=0.35]{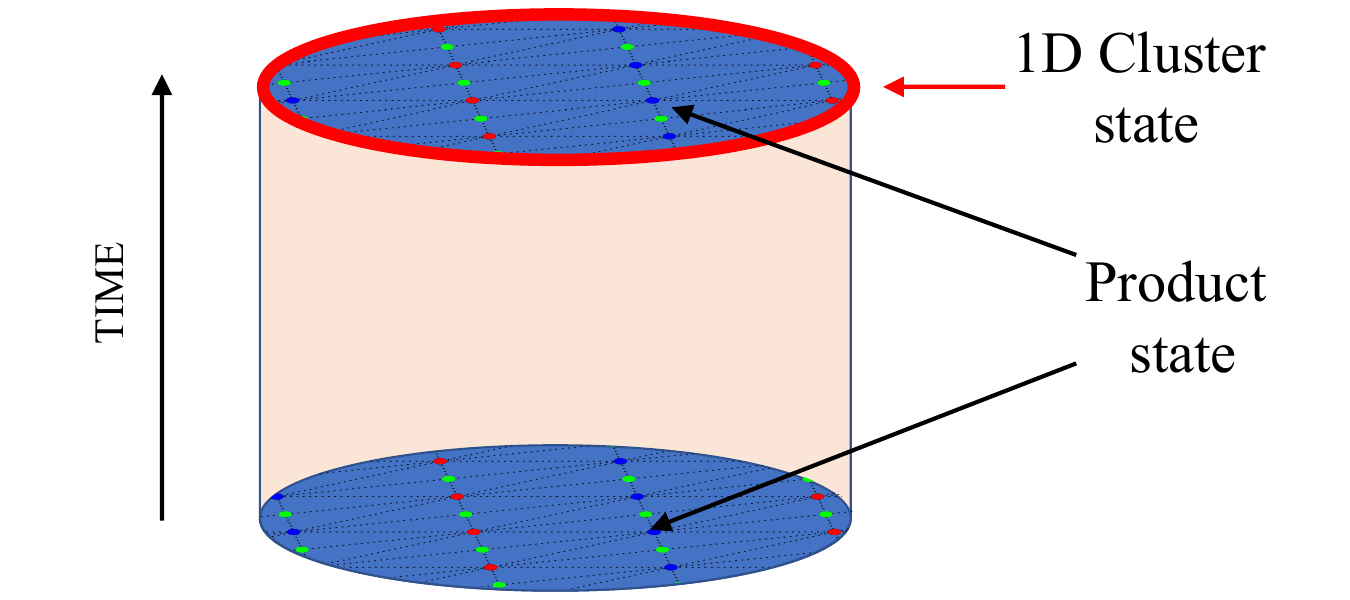}
    \caption{Time evolution by a symmetric three-body Hamiltonian in the 2D bulk pumps a 1D cluster state to the boundary while leaving the bulk invariant. Similarly, 2D cluster states can be prepared at the boundary of a 3D bulk respecting the corresponding subsystem symmetries. In both cases the preparation only takes a finite time, independent of system size. }
    \label{fig:1pump}
\end{figure}

However, there seems to be a drawback for such a convenient property. Although cluster states are easily created by evolving a product state with, for instance, an Ising Hamiltonian for a certain time\cite{BriegelRaussendorf2001}, because the initial and final states belong to different SPT phases, any Hamiltonian evolution that relates the two in finite time necessarily breaks the global(subsystem) symmetry\cite{ChenGuWen2010,HuangChen2015}. Therefore, in experimental setups, unless the Hamiltonian is prepared exactly, the resulting entangled state does not need to be an SPT state, and its use as a resource state is not guaranteed. We seem to come to the conclusion that in order to exploit the universality of the entire SPT phase, one must instead adiabatically prepare the resource state without breaking the symmetry. Such preparation time scales at least linearly in the system size.

In this Letter, we present a method to get around the above argument. Our motivation can be traced back to the seminal work of Thouless \cite{Thouless83}, where an evolution of a 1D system under a symmetric Hamiltonian leaves the bulk invariant after a certain period of time, but can ``pump" quantized amounts of charge from one boundary to another. More recently, higher dimensional generalizations of such a construction have been realized in the field of Floquet SPT phases \cite{ElseNayak2016,vonKeyserlingkSondhi2016,PotterMorimotoVishwanath16,RoyHarper2017,HsinKapustinThorngren20,pivot}, where in fact entire (stationary) phases of matter in one lower dimension can be pumped to the boundary under a finite time evolution while leaving the bulk invariant. Applying this concept, we are able to start with a product state and evolve the system with a Hamiltonian which respects the global(subsystem) symmetry of a 2D(3D) system in such a way that after a fixed finite time --independent of the system size-- a 1D(2D) cluster state is created on the boundary, completely uncoupled from the bulk (Fig. \ref{fig:1pump}). To summarize, the previous no-go argument only holds when the cluster state is assumed to live strictly in the dimension of the defining lattice. Because of additional ancillae coming from the extra dimension of the bulk, the constraint is lifted, and we are able to prepare cluster states both symmetrically and in finite time.

With this setup, we can now take full advantage of the universality of the entire phase. Conceptually, as long as the driving Hamiltonian is modified by any small perturbation that preserves the symmetry, the entangled state on the boundary would still be symmetric and belongs to the same phase as the cluster state. It is therefore still a universal resource state. More realistically, it is possible that symmetric perturbations to the Hamiltonian leave the boundary state coupled to the bulk after the evolution, but we further demonstrate that by performing measurements and feed-forward correction, we can recover a completely decoupled boundary resource state. 

The remainder of the Letter is organized as follows. We first review the notion of cluster states and how they can be viewed as SPT phases. Then, we show how a symmetric 2D Hamiltonian can be used to pump a 1D cluster state to the boundary. This procedure is generalized to pump 2D cluster states to the boundary of a 3D system using a 3D Hamiltonian that respects subsystem symmetry. Lastly, we discuss how to recover a cluster-like resource state in the case that small but symmetric perturbations are added to the Hamiltonian.

\textit{Cluster states.} Let $\ket{0}$ and $\ket{1}$ be $Z$-basis states. Given a graph $\mathcal G=(V,E)$, a graph state\cite{Heinetal2006} is the entangled state
   $\ket{\psi}=  \prod_{ij \in E} CZ_{ij} \bigotimes_{i \in V} \ket{+}_i, $
constructed from initializing with qubits in the $X=1$ eigenstate $\ket{+} \sim \ket{0}+ \ket{1}$ at each vertex and applying the controlled-$Z$ operator $CZ_{ij}  = (-1)^{n_in_j}$, where $n_i = \frac{1- Z_i}{2}$ is the number operator, to every edge of the graph. Equivalently, the graph state is the unique ground state of the stabilizer Hamiltonian $H=-\sum_{i \in V} X_i \prod_{j|(ij) \in E} Z_j$, which is obtained by conjugating $X_i$ on each vertex by the circuit $\prod_{ij \in E}  CZ_{ij}$.

\begin{figure}
\includegraphics[]{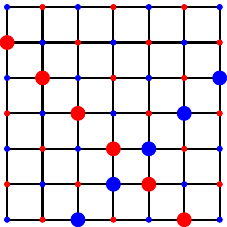}
\caption{The computational power of the 2D cluster state on a square lattice is protected by spin-flip symmetries along individual diagonals lines of the lattice.}
\label{fig:clustersquare}
\end{figure}

\begin{figure}
\includegraphics[]{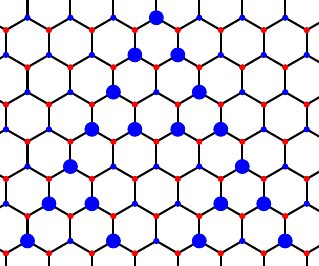}
\caption{The 2D cluster state on the honeycomb lattice. A generator of the fractal symmetries flips all the enlarged blue spins which form a Sierpinski triangle. There is also another set of fractal symmetry generators for the red spins.}
\label{fig:sierpinski}
\end{figure}

When the graph $\mathcal G$ also forms a lattice, the state is called a cluster state. Cluster states are resource states that are universal for MBQC in two or greater spatial dimensions\cite{RaussendorfBriegel2001,VandenNestetal2006}. It was later realized that cluster states are examples of SPT phases\cite{DohertyBartlett2009,ElseBartlettDoherty2012,ElseSchwarzBartlettDoherty12,Youetal2018,Devakuletal2018}. The 1D cluster state is protected by a $\mathbb Z_2^2$ global symmetry, which flips the spins on even and odd sites of the chain respectively. On the other hand, 2D cluster states are SPT phases protected by subsystem symmetry. For example, on a square lattice, the cluster state can be protected by symmetries which flip spins on individual diagonal lines (Fig. \ref{fig:clustersquare}), while on a honeycomb lattice, it can be protected by fractal symmetries which only flip certain spins in the shape of Sierpinski triangles \cite{Devakuletal2018,KubicaYoshida2018} (Fig. \ref{fig:sierpinski}). Furthermore, any state in the same (subsystem) SPT phase as these cluster states (called cluster-like states) can also be used as a universal resource state \cite{Raussendorfetal2017,Stephenetal2017,Raussendorfetal2018,Stephenetal2018,DevakulWilliamson2018}.

\textit{Pumping SPT states protected by Global Symmetries.} To design Hamiltonians whose time evolution pumps SPT states to the boundary, we take inspiration from Floquet SPT phases for bosonic systems with unitary symmetry $G$. The classification of Floquet SPTs protected by $G$ can be thought of as that of a static system with symmetry $G \times \mathbb Z$ where $\mathbb Z$ denotes time translation\cite{ElseNayak2016,vonKeyserlingkSondhi2016,PotterMorimotoVishwanath16}. In the language of group cohomology, we can use the K\"unneth formula to write\footnote{See Appendix B of \cite{ChengZaletelBarkeshliVishwanathBonderson16} for a derivation.}
\begin{equation}
 \mathcal H^{d+1}(G \times \mathbb Z,U(1)) = \mathcal H^{d+1}(G,U(1)) \times \mathcal H^{d}(G,U(1)).
 \end{equation}
The first factor classifies static $G$-SPT phases, while the latter attaches the time-translation symmetry action with $(d-1)$-dimensional $G$-SPTs. It can therefore be interpreted as a drive which pumps such $G$-SPT to the boundary per driving period. We can devise a Hamiltonian to generate this Floquet unitary, which acts as the identity in the bulk, but pumps the SPT phase to the boundary while commuting with the symmetry. The idea is similar to a coupled-layer construction: dividing our $d$-dimensional system (hosting a boundary) into volume-filling ``cells"', the Floquet unitary is obtained by evolving a local Hamiltonian that creates, in one Floquet period, a bubble of the $d-1$-dimensional SPT state along the boundary of each cell. The SPTs cancel in the bulk, leaving only a $(d-1)$-dimensional SPT state on the boundary. Without restrictions to the number of interactions required, the pump for a general bosonic SPT phase can be constructed \cite{RoyHarper2017,Xiong2016}. Here, we build on these works by reducing the weight of the interactions required, focusing on 1+1D SPT phases (since it is undetermined whether there exist SPT phases protected by a global symmetry in higher dimensions whose entire phase is universal). Subsequently, we will turn to pumps for subsystem SPT phases in higher dimensions, which are new, and for which a formal classification has not been put forward.

\begin{figure}
\includegraphics[]{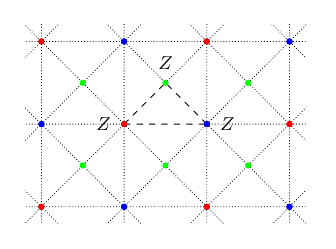}
\caption{The Union-Jack lattice, with a global $\mathbb Z_2^2$ symmetry defined as flipping spins on two of the three colors. The Hintermann-Merlini 3-body interaction Eq.\eqref{equ:HMHam} commutes with this symmetry.}
\label{fig:3bodyunionjack}
\end{figure}

\textit{Pumping the 1D cluster state.}
Let us demonstrate how to prepare the 1D cluster state on the boundary of the Union-Jack lattice respecting a global $\mathbb Z_2^2$ symmetry using only three-body interactions, an improvement over previous setups using four-body interactions\cite{PotterMorimoto2017,RoyHarper2017}\footnote{this is reviewed in the Supplementary Material which includes Ref.~\cite{BaxterWu73}.}. We place qubits on the vertices on the Union-Jack lattice, which is three-colorable as red, blue, and green as shown in Fig. \ref{fig:3bodyunionjack}. The global $\mathbb Z_2^2$ symmetry is defined via the action of its three $\mathbb Z_2$ subgroups, which flip spins in the $Z$-basis on two of the three colors. Starting with the product state $\bigotimes_{v\in V} \ket{+}_v$, we will evolve our system with the following Hintermann-Merlini Hamiltonian \cite{HintermannMerlini72} for time $\pi/4$
\begin{align}
    H = -\sum_{\Delta_{123}} Z_1 Z_2 Z_3,
    \label{equ:HMHam}
\end{align}
where the sum is over triangles $\Delta_{123}$ of all orientations. This Hamiltonian commutes with the $\mathbb Z_2^2$ symmetry. 

To see the action of the resulting unitary, we expand using $Z_i = 1-2n_i$, and find
\begin{align}
 \exp -i\frac{\pi}{4} Z_1Z_2Z_3  &=  e^{-i\frac{\pi}{4}}e^{i\frac{\pi}{2}(n_1+n_2+n_3)}e^{i\pi (n_1n_2 +n_2n_3+n_3n_1)} \nonumber\\
 & \propto S_1S_2S_3CZ_{12}CZ_{23}CZ_{31} \label{equ:calculation1D}
\end{align}
Hence, the local three-body term exponentiates to a product of $S = e^{\frac{\pi i}{2} n}$ gates for each vertex and $CZ$ gates for each edge of the triangle. Taking the product of such local unitaries for all triangles, each vertex is always acted by either four or eight $S$ gates, which cancel both in the bulk and on the boundary. On the other hand, the $CZ$ gates cancel pairwise in the bulk, leaving (up to an overall phase) $\exp -iH\frac{\pi}{4} \propto \prod_{ij \in \partial M}  CZ_{ij}$, where $\partial M$ denotes the boundary spins of the lattice. Therefore this unitary pumps the cluster state to the boundary.

\begin{figure}[t!]
\includegraphics[scale=0.6]{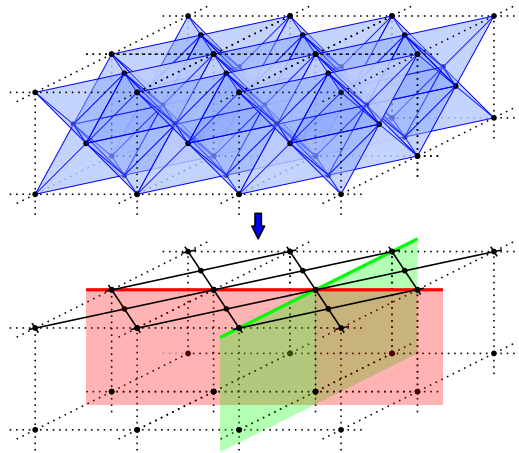}
\caption{A unitary evolution of the tetrahedral Ising interaction on the FCC lattice pumps the 2D cluster state to the entire boundary (shown here for the $(001)$ boundary). The local unitary (blue tetrahedron) is generated by a four-body term in Eq. \eqref{equ:tetdef}. Planar symmetries (red and green) terminate on the boundary as line symmetries. }
\label{fig:tetrahedron}
\end{figure}

\textit{Pumping subsystem SPTs.} We will now generalize the results to prepare 2D cluster states. Note that such states can already be prepared strictly in 2D in the presence of only global symmetries. Nevertheless, they belong to the trivial phase under such symmetries and thus we cannot exploit the universality of the phase. Thus, we need the presence of subsystem symmetries, which requires a 3D pump.

First, consider the FCC lattice with planar subsystem symmetries defined as flipping spins in individual $(100)$, $(010)$ or $(001)$ planes. These planar symmetries terminate as line symmetries on the boundary. Our driving Hamiltonian will be a four-body tetrahedral Ising interaction\cite{VijayHaahFu2016}
\begin{equation}
H = -\sum_{\tetrahedron_{1234}} Z_1Z_2Z_3Z_4
\end{equation}
where each tetrahedron consists of a vertex along with three adjacent face centers within the same cube. This Hamiltonian commutes with the planar symmetries. Evolving the product state with the above Hamiltonian for time $\pi/4$, a similar calculation to Eq. \eqref{equ:calculation1D} shows that visually,
\begin{align}
\raisebox{-0.5\height}{\includegraphics[scale=0.7]{tet.pdf}} \equiv \exp \left[-i\frac{\pi}{4}  \raisebox{-0.5\height}{\includegraphics[scale=0.7]{ZZZZ.pdf}}\right]= \raisebox{-0.5\height}{\includegraphics[scale=0.7]{SCZ.pdf}}
\label{equ:tetdef}
\end{align}
where each dense edge of the tetrahedron denotes a $CZ$ gate. Taking the product over all tetrahedra in the bulk (Fig. \ref{fig:tetrahedron}), we are left with $CZ$ gates acting only along the boundary. The cluster state on the (rotated) square lattice (Fig. \ref{fig:clustersquare}) can therefore be prepared on the $(100)$, $(010)$ and $(001)$ boundaries.  In the SM, we show how to prepare the triangular lattice cluster state by instead choosing the $(111)$ boundary, and an alternative method to prepare it on the boundary of the cubic lattice.

For our second example, we will prepare the 2D cluster state on the honeycomb lattice. Our 3D bulk is a stack of 2D honeycombs with the two sites per unit cell labeled red and blue, as in Fig. \ref{fig:sierpinski}. The fractal symmetry is defined as acting the fractal symmetry action of Fig. \ref{fig:sierpinski} simultaneously for every layer. Consider the following gates defined for each blue and red vertex, respectively
\begin{align*}
V_{v_b} &= \raisebox{-.5\height}{ \includegraphics[]{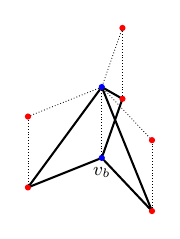} }  
, &
V_{v_r}& = \raisebox{-.5\height}{ \includegraphics[]{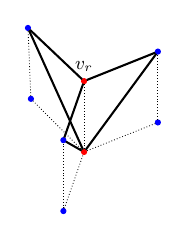} } 
\end{align*}
where each solid line denotes a $CZ$ gate. This can be expanded using $CZ_{ij} = \frac{1}{2}(1+Z_i + Z_j - Z_iZ_j)$ to a sum of at most five $Z$ operators. The blue fractal symmetries trivially commute with $V_{v_b}$, while the red fractal symmetries around any blue site only flips zero or two of the three adjacent red sites within in each layer. Therefore, $V_{v_b}$ commutes with the all the fractal symmetries and similarly for $V_{v_r}$. The product $\prod_{v} V_{v_b} V_{v_r} = e^{-i\frac{\pi}{2}H}$ over all vertices $v$ in the 3D lattice creates two cluster states on the top and bottommost honeycomb layers, where
$H= \sum_v V_{v_b} + V_{v_r}$.

 Generalizing this, it is possible to similarly prepare any 2D fractal cluster state\cite{Devakuletal2018} generated by some 1D cellular automaton. Here, we will give the underlying argument, and prove it rigorously in the SM\footnote{See Supplementary Material, which includes Refs.~\cite{Haah2013,NewmanMoore1999,ChenLuVishwanath2014}}. For each blue site $v_b$, define $V_{v_b}$ to be a product of $CZ$ operators connecting $v_b$ and the blue site directly above it to its nearest neighbor red sites. Any symmetry generated by the cellular automaton will only flip an even number of the nearest neighbor red sites, so these gates are symmetric. Analogously, for each red site $v_r$, $V_{v_r}$ is a product of $CZ$ operators connecting  $v_r$ and the red site directly below it to its nearest neighbor blue sites. A product of such gates over all vertices creates the cluster state at the top and bottom-most layers, so the sum of these gates is exactly our desired driving Hamiltonian.

\textit{Recovering cluster-like states in practical setups.} Finally, we discuss how to take into account possible undesirable perturbations that could be introduced into the driving Hamiltonian when implemented in practice. These perturbations could entangle the boundary state with the bulk, rendering it useless as a resource state. However, we will show that as long as these perturbations are small and respect the symmetry, measurements in the bulk followed by feed-forward correction can recover a resource state in the same phase as the cluster state. 

The basic idea is as follows. Suppose the driving Hamiltonian is perturbed by symmetric local terms, whose operator norms are bounded above by  $\epsilon ||H||$ for some small $\epsilon$. To prepare the resource state, we choose a bulk which is much larger than the support of possible perturbations and initialize all qubits to the all $\ket{+}$ state. We now consider how the terms possibly affect the cluster state on the boundary after the evolution.
\begin{enumerate}
    \item If the perturbation acts purely in the bulk, then our resource state on the boundary is not affected.
    \item If the perturbation acts purely on the boundary, then the state is perturbed symmetrically, which will still be a valid resource state as long as $\epsilon$ is small enough to keep it in the SPT phase.
    \item If the perturbation acts both in the bulk and on the boundary, then this term could break the symmetry restricted to the boundary or bulk separately, while preserving the total symmetry of whole system. In that case, the term will flip an odd number of $\ket{+}$ states in the bulk to $\ket{-}$. Therefore, we can eliminate this error by performing a measurement in the $X$-basis for all qubits in the bulk, and if we measure an odd number of $\ket{-}$ states along any bulk symmetry operator, we apply single-spin flips $X$ on the boundary to recover a symmetric state. 
\end{enumerate}

\textit{Discussion.} Inspired by quantum pumps and Floquet SPT phases, we devised a 2D(3D) Hamiltonian which respects the global(subsystem) symmetry extended into the bulk, and showed that a product state driven by this Hamiltonian for a fixed time independent of system size prepares a 1D(2D) cluster state on the boundary. Then, exploiting the universality of the entire symmetry-protected phase, we were able to guarantee the preparation of a resource state even when the Hamiltonian is not implemented exactly as long as perturbations are small and symmetric. This was achieved by followup measurements and feed-forward correction. We find it remarkable that topology proves itself useful in methods beyond topological quantum computing.

We conclude with prospects for future work.

From the point of view of topological phases, our results entails that intrinsically interacting Floquet SPT phases protected by subsystem symmetry are at least classified by subsystem SPT phases in one lower dimension, identical to the global symmetry case. It would be interesting to see whether this classification is complete. Furthermore, gauging Floquet subsystem SPT phases can give rise to Floquet fracton orders, where gapped excitations with restricted mobility are dynamically enriched into non-abelian excitations via the Floquet drive\cite{PotterMorimoto2017}. There is also an intriguing connection between pumps and transversal logical gates of the gauged topological codes that deserve exploration in the case of subsystem symmetries\footnote{See Supplementary Material, which includes Refs.~\cite{CobaneraOrtizNussinov2011,Yoshida2017,Pretko2018,ShirleySlagleChen2019,Radicevic2019,KubicaBeverland15,Yoshida2015,Yoshida2017,Kesselringetal18,WebsterBartlett18}}

For future prospects for MBQC,  
we have presented three(four)-body interactions to symmetrically prepare one(two)-dimensional cluster states. It would be interesting if this number can be further lowered given that universal resource states can arise as ground states of two-body Hamiltonians\cite{Chenetal2009,WeiAffleckRaussendorf2011,WeiAffleckRaussendorf2012}. In addition, current computational schemes implicitly assume the cluster-like states possess translation-invariance\cite{ElseSchwarzBartlettDoherty12,Raussendorfetal2017,Raussendorfetal2018,DevakulWilliamson2018,DanielAlexanderMiyake20}\footnote{More specifically, that the state can be written as a translation-invariant tensor network}, which might not hold for the states prepared using this method. It would be crucial to devise a computational scheme which relaxes such assumption. Finally, we hope to investigate whether there are experimental platforms where such global or subsystem symmetries are inherent or arise as an approximate symmetry.  Ultimately, finding a Hamiltonian that can be faithfully implemented experimentally, and a way to limit perturbations to ones that respect the symmetry,  would provide a scalable and reliable method to create universal resource states.
\begin{acknowledgments}
\textit{Achknowledgments} We thank Yanzhu Chen, Ruihua Fan, Dominic Else, Andrew Potter, Abhishodh Prakash, Robert Raussendorf, Ryan Thorngren, Ruben Verresen, Sagar Vijay, and Tzu-Chieh Wei for helpful discussions. We also thank the referees for suggestions that were crucial to improving the manuscript. NT is grateful to the Yukawa Institute for Theoretical Physics at Kyoto University. Discussions during the workshop YITP-T-19-03 ``Quantum Information and String Theory 2019" were useful to complete this work. NT acknowledges the support of NSERC. AV was supported by the DARPA DRINQS program (award D18AC00033) and by Simons Investigator award (AV) and by the Simons Collaboration on Ultra-Quantum Matter, which is a grant from the Simons Foundation (651440, AV).
\end{acknowledgments}

\bibliography{references}
\onecolumngrid
\clearpage
\twocolumngrid

\end{document}


\tikzset{
    vertex/.style={fill,circle,draw,scale=0.3}
}

\newcommand{\tetrahedron}{
  \mathchoice
    {\includegraphics[height=1ex]{tetrahedron}} 
    {\includegraphics[height=1ex]{tetrahedron}} 
    {\includegraphics[height=1.5ex]{tetrahedron}} 
    {\includegraphics[height=.5ex]{tetrahedron}} 
}


\def\bra#1{\mathinner{\langle{#1}|}}
\def\ket#1{\mathinner{|{#1}\rangle}}
\newcommand{\Ket}[1]{\vcenter{\hbox{$\displaystyle\stretchleftright{|}{#1}{\bigg\rangle}$}}}

\def\cube{\raisebox{-.5\height}{\begin{tikzpicture}[scale=0.1]
\def\xone{-0.9};
\def\xtwo{-1.4};
\def\yone{2};
\def\ytwo{0};
\def\zone{0};
\def\ztwo{2};
\coordinate (000)      at (0,0) {};
 \coordinate (100)         at (\xone,\xtwo) {};
 \coordinate (010)      at (\yone,\ytwo) {};
 \coordinate (001)      at (\zone,\ztwo) {};
 \coordinate(101)       at (\xone+\zone,\xtwo+\ztwo) {};
 \coordinate (110)      at (\xone+\yone,\xtwo+\ytwo) {};
\coordinate (111)      at (\xone+\yone+\zone,\xtwo+\ytwo+\ztwo) {};
\coordinate  (011)         at (\yone+\zone,\ytwo+\ztwo) {};
\draw[-,color=gray] (000) -- (100) {};
\draw[-,color=gray] (001) -- (101) {};
\draw[-,color=gray] (100) -- (101) {};
\draw[-,color=gray] (000) -- (001) {};
\draw[-,color=gray] (011) -- (111) {};
\draw[-,color=gray] (110) -- (111) {};
\draw[-,color=gray] (100) -- (110) {};
\draw[-,color=gray] (101) -- (111) {};
\draw[-,color=gray] (001) -- (011) {};
\draw[-,color=gray] (000) -- (010) {};
\draw[-,color=gray] (010) -- (110) {};
\draw[-,color=gray] (010) -- (011) {};
\end{tikzpicture}}}

\title{Supplementary Material\\for\\
Symmetric Finite-Time Preparation of Cluster States via Quantum Pumps}

\author{Nathanan Tantivasadakarn}
\affiliation{Department of Physics, Harvard University, Cambridge, MA 02138, USA}
\author{Ashvin Vishwanath}
\affiliation{Department of Physics, Harvard University, Cambridge, MA 02138, USA}

\maketitle

\section{Pumping the 1D Cluster state on the triangular and square lattices}\label{app:1Dsquarelattice}

\begin{figure}[h!]
\includegraphics[]{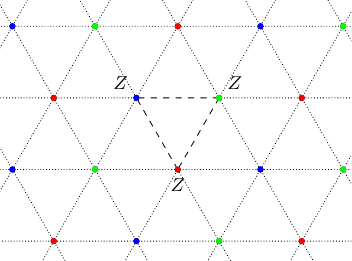}
\caption{Triangular Lattice, with $\mathbb Z_2^2$ symmetry defined as flipping spins on two of the three colors. The Baxter-Wu 3-body interaction commutes with this symmetry.}
\label{fig:3body}
\end{figure}
We explain a setup to pump cluster states symmetrically to the boundary of the triangular and square lattices. The setup for the triangular lattice is very similar to that of the Union Jack lattice given in the main text. Qubits are placed on the vertices of the triangular lattice, which are 3-colorable as red, blue, and green as shown in Fig. \ref{fig:3body}. The global $\mathbb Z_2^2$ symmetry is defined via the action of its three $\mathbb Z_2$ subgroups, which flip spins in the computational basis on two of the three colors. Starting with a product state, we will evolve our system with the following Baxter-Wu Hamiltonian \cite{BaxterWu73} for time $\pi/4$
\begin{align}
    H = -\sum_{\Delta_{123}} Z_1 Z_2 Z_3,
    \label{equ:BWHam}
\end{align}
where the sum is over all up and down triangles $\Delta_{123}$ in the triangular lattice. This Hamiltonian commutes with the $\mathbb Z_2^2$ symmetry. 

The unitary obtained from evolving the Hamiltonian up to an overall phase is
\begin{align}
U &= \exp -iH\frac{\pi}{4} \propto \prod_{ij \in \partial M} Z_i CZ_{ij} \prod_{i \in M} Z_i,
\end{align}
where $M$ and $\partial M$ are the bulk and boundary spins of the lattice, respectively. Therefore this unitary pumps a cluster-like state to the boundary
and flips all $\ket{+}$ states in the bulk to $\ket{-}$. Because of this action in the bulk, we must instead post-select an even number of $\ket{+}$ states in the presence of symmetric perturbations. Alternatively, the action of $Z$-operators in the bulk can be removed by ``staggering" the Baxter-Wu Hamiltonian. That is, flipping the sign of the Hamiltonian for all upside-down triangles. In this case one finds
\begin{align}
U &= \exp -iH\frac{\pi}{4} \propto \prod_{ij \in \partial M} CZ_{ij} \prod_{i \in M} Z_i,
\end{align}
which exactly prepares the cluster state on the boundary, while leaving all $\ket{+}$ states invariant in the bulk.

\begin{figure}[t!]
\includegraphics[]{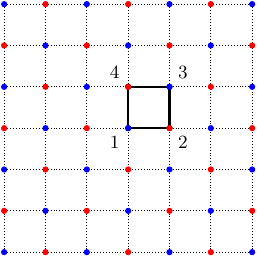}
\caption{Two $\mathbb Z_2$ symmetries on a square lattice in 2D act by flipping all the red and blue spins respectively. The driving Hamiltonian obtained by summing over product four $CZ$ gates at the edge of all squares pumps a 1D cluster state to the boundary.}
\label{fig:1Dpump}
\end{figure}
The setup for the square lattice was originally discussed in \cite{PotterMorimoto2017} in the context of Floquet SPTs and requires four-body interactions. We endow a lattice of spins in 2D with a $\mathbb Z_2^2$ symmetry which acts as spin flip on either the red or blue sites, respectively as shown in Figure \ref{fig:1Dpump}. The qubits are all initialized in the $\ket{+}$ state. Now, for each square $\Box_{1234}$ in the lattice, consider the following Hamiltonian, which is a sum over a product of four $CZ$ operators around each square.
\begin{align}
\label{equ:1DclusterHam}
H &= \sum_{\Box_{1234}} CZ_{12} CZ_{23} CZ_{34}CZ_{41} \nonumber \\
  &= \frac{1}{2} \sum_{\Box_{1234}} (1 + Z_1Z_3 + Z_2Z_4 -Z_1Z_2Z_3Z_4)
\end{align}
This Hamiltonian commutes with the $\mathbb Z_2^2$ symmetry. Next, if we evolve the product state using the above Hamiltonian for time $\pi/2$, the resulting unitary evolution up to a phase is
\begin{align}
U &= \exp -iH\frac{\pi}{2} \nonumber \\
&= \prod_{\Box_{1234}} \left [\cos \frac{\pi}{2} - i \sin \frac{\pi}{2} CZ_{12} CZ_{23} CZ_{34}CZ_{41}  \right] \nonumber \\
&\propto \prod_{ij \in \partial M} CZ_{ij},
\end{align}
where $\partial M$ are qubits on the boundary of the square lattice. Here, we used the fact that all the terms in the Hamiltonian mutually commute and all square to the identity. We see that the resulting unitary acts trivially in the bulk, since all the $CZ$ operators cancel pairwise, and so we are left with $CZ$'s acting only along the boundary. Thus, the unitary creates a 1D cluster state along the boundary.

\begin{figure}[t!]
\includegraphics[scale=0.8]{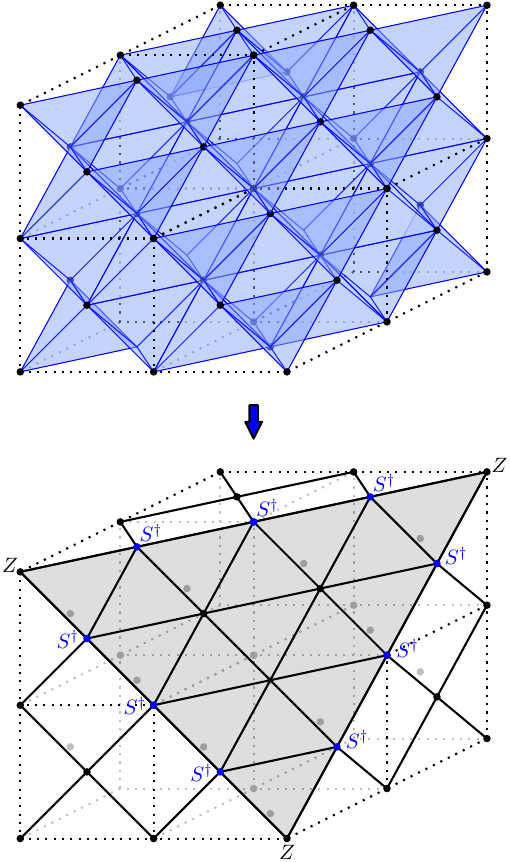}
\caption{The $(111)$ boundary of the FCC lattice (shown in grey as a vertex figure of one of the corners of the cube) can host a cluster state on the triangular lattice, which is an SSPT protected by three directions of line subsystem symmetries. The residual phase gates shown are residuals of the preparation and only exists on the interface between different boundaries. They are absent in the ``bulk" of the 2D SSPT.}
\label{fig:111pump}
\end{figure}
\section{Pumping the 2D cluster state on the triangular lattice}\label{}
We show how to prepare the 2D cluster state on the triangular lattice using two methods. The first is to use the same 3D Hamiltonian as the main text, but instead expose the $(111)$ boundary of the FCC lattice. This cluster state is an SPT protected by three directions of line subsystem symmetries, defined as the termination of the three planar symmetries on the $(111)$ boundary. This is shown in Fig. \ref{fig:111pump}.

The second method is to use the usual cubic lattice, with planar subsystem symmetries flipping spins on the $(001)$, $(010)$, and $(100)$ planes. We will consider the$(001)$, $(010)$, and $(100)$ boundary terminations, where two of the planar subsystem symmetries terminate as line subsystem symmetries on the boundary. We note that the cluster state on the triangular lattice is still protected even if it is protected by only two of the three directions of line symmetries\cite{DevakulWilliamsonYou2018,TantivasadakarnVijay2019}. This is the state that will be prepared on the boundary.

\begin{figure}[t!]
\includegraphics[scale=0.7]{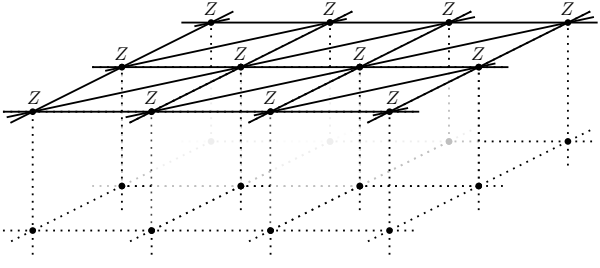}
\caption{An evolution of the Hamiltonian \eqref{equ:Hamiltoniancube} for time $\pi/4$ pumps a cluster-like state on the triangular lattice to the boundary of the cubic lattice.}
\label{fig:triangularpump}
\end{figure}

The Hamiltonian we will use is a sum of the following four-body interactions over all cubes
\begin{align}
    H = -\sum_{\cube}& \left [
\raisebox{-.5\height}{\begin{tikzpicture}[scale=0.7]
\def\xone{-0.9};
\def\xtwo{-1.4};
\def\yone{2};
\def\ytwo{0};
\def\zone{0};
\def\ztwo{2};
\coordinate (000)      at (0,0) {};
 \coordinate (100)         at (\xone,\xtwo) {};
\node[label=center:$Z$] (010)      at (\yone,\ytwo) {};
\node[label=center:$Z$] (001)      at (\zone,\ztwo) {};
\node[label=center:$Z$](101)       at (\xone+\zone,\xtwo+\ztwo) {};
\node[label=center:$Z$] (110)      at (\xone+\yone,\xtwo+\ytwo) {};
\coordinate (111)      at (\xone+\yone+\zone,\xtwo+\ytwo+\ztwo) {};
\coordinate  (011)         at (\yone+\zone,\ytwo+\ztwo) {};
\draw[-,densely dotted,color=gray] (000) -- (100) {};
\draw[-,densely dotted,color=gray] (001) -- (101) {};
\draw[-,densely dotted,color=gray] (100) -- (101) {};
\draw[-,densely dotted,color=gray] (000) -- (001) {};
\draw[-,densely dotted,color=gray] (011) -- (111) {};
\draw[-,densely dotted,color=gray] (110) -- (111) {};
\draw[-,densely dotted,color=gray] (100) -- (110) {};
\draw[-,densely dotted,color=gray] (101) -- (111) {};
\draw[-,densely dotted,color=gray] (001) -- (011) {};
\draw[-,densely dotted,color=gray] (000) -- (010) {};
\draw[-,densely dotted,color=gray] (010) -- (110) {};
\draw[-,densely dotted,color=gray] (010) -- (011) {};
\end{tikzpicture}} +
\raisebox{-.5\height}{\begin{tikzpicture}[scale=0.7]
\def\xone{-0.9};
\def\xtwo{-1.4};
\def\yone{2};
\def\ytwo{0};
\def\zone{0};
\def\ztwo{2};
\coordinate (000)      at (0,0) {};
\node[label=center:$Z$] (100)      at (\xone,\xtwo) {};
\node[label=center:$Z$] (010)      at (\yone,\ytwo) {};
\coordinate (001)      at (\zone,\ztwo) {};
\node[label=center:$Z$] (101)      at (\xone+\zone,\xtwo+\ztwo) {};
\coordinate (110)      at (\xone+\yone,\xtwo+\ytwo) {};
\coordinate (111)      at (\xone+\yone+\zone,\xtwo+\ytwo+\ztwo) {};
\node[label=center:$Z$]  (011)         at (\yone+\zone,\ytwo+\ztwo) {};
\draw[-,densely dotted,color=gray] (000) -- (100) {};
\draw[-,densely dotted,color=gray] (001) -- (101) {};
\draw[-,densely dotted,color=gray] (100) -- (101) {};
\draw[-,densely dotted,color=gray] (000) -- (001) {};
\draw[-,densely dotted,color=gray] (011) -- (111) {};
\draw[-,densely dotted,color=gray] (110) -- (111) {};
\draw[-,densely dotted,color=gray] (100) -- (110) {};
\draw[-,densely dotted,color=gray] (101) -- (111) {};
\draw[-,densely dotted,color=gray] (001) -- (011) {};
\draw[-,densely dotted,color=gray] (000) -- (010) {};
\draw[-,densely dotted,color=gray] (010) -- (110) {};
\draw[-,densely dotted,color=gray] (010) -- (011) {};
\end{tikzpicture}} \right. \nonumber\\
&+\left. \raisebox{-.5\height}{\begin{tikzpicture}[scale=0.7]
\def\xone{-0.9};
\def\xtwo{-1.4};
\def\yone{2};
\def\ytwo{0};
\def\zone{0};
\def\ztwo{2};
\coordinate (000)      at (0,0) {};
\node[label=center:$Z$] (100)      at (\xone,\xtwo) {};
\coordinate  (010)      at (\yone,\ytwo) {};
\node[label=center:$Z$](001)      at (\zone,\ztwo) {};
\coordinate (101)      at (\xone+\zone,\xtwo+\ztwo) {};
\node[label=center:$Z$] (110)      at (\xone+\yone,\xtwo+\ytwo) {};
\coordinate (111)      at (\xone+\yone+\zone,\xtwo+\ytwo+\ztwo) {};
\node[label=center:$Z$]  (011)         at (\yone+\zone,\ytwo+\ztwo) {};
\draw[-,densely dotted,color=gray] (000) -- (100) {};
\draw[-,densely dotted,color=gray] (001) -- (101) {};
\draw[-,densely dotted,color=gray] (100) -- (101) {};
\draw[-,densely dotted,color=gray] (000) -- (001) {};
\draw[-,densely dotted,color=gray] (011) -- (111) {};
\draw[-,densely dotted,color=gray] (110) -- (111) {};
\draw[-,densely dotted,color=gray] (100) -- (110) {};
\draw[-,densely dotted,color=gray] (101) -- (111) {};
\draw[-,densely dotted,color=gray] (001) -- (011) {};
\draw[-,densely dotted,color=gray] (000) -- (010) {};
\draw[-,densely dotted,color=gray] (010) -- (110) {};
\draw[-,densely dotted,color=gray] (010) -- (011) {};
\end{tikzpicture}} \right].
\label{equ:Hamiltoniancube}
\end{align}
 This Hamiltonian commutes with the planar symmetry. The unitary obtained by evolving this Hamiltonian by $\pi/4$ up to a phase is
\begin{equation}
U = e^{-i \frac{\pi}{4}H} \propto \prod_{\cube} \raisebox{-0.5\height}{  \includegraphics[]{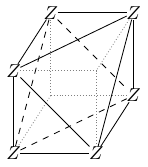}}
\end{equation}
where the dense an dashed lines denote $CZ$ gates. These local operators cancel in the bulk, but creates the 2D cluster-like state\footnote{which differs from the actual cluster state by an application of Pauli-$Z$'s on all boundary sites} on the triangular lattice on all three boundaries. This is shown for the $(001)$ boundary in Fig. \ref{fig:triangularpump}.

\section{Cellular Automata and Fractal SPTs}
We summarize some important notions of Cellular Automata (CA), fractal symmetries, and construction of fractal SPTs. We will then show how to disentangle two copies of any fractal SPT and hence write down a driving Hamiltonian that pumps fractal SPTs to the boundary. 

\subsection{Algebraic Formalism}

The notation used here closely follows that of Ref. \onlinecite{Devakuletal2018}. For a formal treatment, see for example Ref. \onlinecite{Haah2013}.
The positions of qubits on a 2D lattice can be parametrized by coordinates $(i,j) \in \mathbb Z^2$. The Hamiltonian can be written as a sum of stabilizers where each term is product of $X$ and $Z$ operators. The location of the Paulis can be given by a Laurent polynomial $\alpha \in \mathbb F_2[x,\bar x,y,\bar y]$, where $\bar x = x^{-1}$ and $\bar y = y^{-1}$. Since the expansion
\begin{equation}
\alpha = \sum_{i,j=-\infty}^\infty \alpha_{ij} x^iy^j.
\end{equation}
for $\alpha_{ij} \in \mathbb F_2$ is unique, the operators
\begin{equation}
Z(\alpha) = \prod_{ij} Z_{ij}^{\alpha_{ij}},\\
X(\alpha) = \prod_{ij} X_{ij}^{\alpha_{ij}},
\end{equation}
places $Z$ or $X$ operators at sites $(i,j)$ whenever $\alpha_{ij}=1$. In this notation, translation by the vector $(i,j)$ is realized by multiplication of the monomial $x^iy^j$, respectively.

Given two polynomials $\alpha$ and $\beta$, their commutation polynomial $P(\alpha,\beta)$ is defined as
\begin{equation}
P(\alpha,\beta)=\alpha \bar \beta= \sum_{ij} P_{ij}(\alpha,\beta)x^iy^j.
\end{equation}
The operators $X(\alpha)$ and $Z(\beta)$ can anticommute if there is an odd number of $X$ and $Z$ operators appearing at the same position. This corresponds to the number of terms present in both $\alpha$ and $\beta$. Thus, we can calculate $\alpha \bar \beta$ and look at the $x^0 y^0$ term. 
In other words, $X(\alpha)$ and $Z(\beta)$ will anticommute iff  $P_{00}(\alpha,\beta)=1$ and will commute iff $P_{00}(\alpha,\beta)=0$.

In general, we can shift $Z$ by $x^iy^j$ and look at the commutation between  $X(\alpha)$ and $Z(x^iy^j\beta)$. The commutation is determined by the $x^0y^0$ coefficient of $\alpha \overline{x^iy^j \beta}$ or equivalently, the $x^iy^j$ coefficent of $\alpha \bar \beta$, which is $P_{ij}(\alpha,\beta)$. Hence, we see that
\begin{equation}
 X(\alpha)Z(x^iy^j\beta) = (-1)^{P_{ij}(\alpha,\beta)}Z(x^iy^j\beta)X(\alpha).
\end{equation}

We also find it useful to extend the algebraic notation to $CZ$ operators. For simplicity, let us assume that $CZ$ always acts between two different sublattices, which we distinguish by the presence of a monomial $s$. We can demand that the input of $CZ$ is always of the form 
$ CZ(\beta,\gamma s)$, where $\beta=\sum_{ij}\beta_{ij} x^iy^j$, and $\gamma=\sum_{ij}\gamma_{ij} x^iy^j$. Then, we define
\begin{equation}
 CZ(\beta,\gamma s) = \prod_{ijkl}CZ_{ij,kl}^{\beta_{ij}\gamma_{kl}}.
 \end{equation}
That is, it acts $CZ$ on all possible combinations between sites with non-zero $\beta_{ij}$ sites in the $s^0$ sublattice and non-zero $\gamma_{ij}$ sites in the $s^1$ sublattice.

Now, if we conjugate $X(\alpha)$ with $CZ(\beta,\gamma s)$, this creates $Z$ operators at $\gamma s$ depending on the number of overlaps between $\alpha$ and $\beta$. Again, this can be expressed in terms of the commutation polynomial
\begin{equation}
CZ(\beta,\gamma s)X(\alpha)CZ(\beta, \gamma s) = X(\alpha) Z(\gamma s)^{P_{00}(\alpha,\beta)},
\end{equation}
and in general,
\begin{equation}
CZ(x^iy^jz^k\beta,\gamma s)X(\alpha)CZ(x^iy^jz^k\beta, \gamma s) = X(\alpha) Z(\gamma s)^{P_{ijk}(\alpha,\beta)}.
\label{equ:CZcomm1}
\end{equation}
Swapping roles of the $s^0$ and $s^1$ sublattice, we can also obtain
\begin{equation}
CZ(\beta,x^iy^jz^k \gamma s)X(\alpha s)CZ(\beta, x^iy^jz^k \gamma s) = X(\alpha s) Z(\beta)^{P_{ijk}(\alpha,\gamma)}.
\label{equ:CZcomm2}
\end{equation}

\subsection{Fractal Symmetries}
A cellular automaton (CA) can be generated from a function $f \in \mathbb F_2[x,\bar x]$. With such a function, we can construct the Hamiltonian
$$ H= - \sum_{ij} Z(x^i y^j (1+ \bar f \bar y)) $$
i.e. the Hamiltonian is a product of $Z$ at coordinate $i,j$ along with $Z$ operators at position $j-1$ given according to $\bar f$. As an example, the Sierpinski rule $f=1+x$ gives 
$$ H= - \sum_{ij} Z_{i-1,j-1}Z_{i,j-1}Z_{i,j} $$
which is the Newman-Moore Hamiltonian\cite{NewmanMoore1999}.

The symmetry of this Hamiltonian can be generated by any function $q(x) \in \mathbb F_2[x]$ as $$S(q(x)) = X\left( q(x) \mathcal F(x,y) \right)$$
where
\begin{equation}
\mathcal F(x,y)=\sum_{t=0}^\infty f(x)^ty^t.
\label{equ:mathcalF}
\end{equation}
Intuitively, the Pauli $X$ operators on the first row $(y=0)$ are given by $q(x)$ and subsequent rows are obtained from the previous row using the update rule $f(x)$. For example, choosing $q(x)=x^j$ flips spins on a Sierpinski triangle with apex at site $j$ on the first row (see Fig. 3 of the main text without the red sites for a visualization).

Suppose our Hamiltonian is defined in a region where $y \ge 0$ (whether $x$ is be unbounded or periodic does not affect the argument.) To prove that the symmetry defined commutes with the Hamiltonian, we need to check that the coefficients $P_{ij}(\alpha,\beta)$ of the commutation polynomial with $\alpha = q(x)\sum_{t=0}^\infty f(x)^ty^t$ and $\beta = 1+\bar f \bar y$ vanishes for all $j>0$ (because there are no terms in the Hamiltonian for $j \le 0$). Computing, we find
\begin{align}
P(\alpha,\beta) &= q(x) (1+f y)\sum_{t=0}^\infty f(x)^ty^t =q(x).
\end{align}
Therefore, $P_{ij}$ vanishes for all $j>0$.

\subsection{Fractal SPTs}
Let us now construct an SPT protected by fractal symmetries. Starting with a product state Hamiltonian containing two sublattices (labeled by $s^0$ and $s^1$) per site
\begin{equation}
H_0 = - \sum_{ij} \left [X(x^i y^j) + X(x^i y^j s)  \right].
\end{equation}
Given a  CA $f$, the Hamiltonian has symmetries
\begin{align}
S(q(x)) &= X(q(x) \mathcal F(x,y)  ),  \\
S'(q(x)) &= X(q(x) \bar{\mathcal F}(x,y) s ),
\end{align}
with $\mathcal F$ given by Eq. \ref{equ:mathcalF}. We now evolve $H_0$ with a unitary
\begin{align}
 U &= \prod_{ij} CZ \left ( x^iy^j (1+ \bar f \bar y), x^iy^js   \right) \nonumber\\
 &= \prod_{ij} CZ \left (x^iy^j ,x^iy^j (1+  f  y) s \right)
\end{align}
The two expressions of the unitary above can be shown to be equivalent by expanding $\bar f = \sum_i f_i \bar x^i$ and performing the appropriate shifting of indices in the product. Since the unitary contains only $CZ$ operators, the ground state of this Hamiltonian is a cluster state.  We remark that this can be thought of decorating the charge excitations one sublattice to the symmetry defects of the other sublattice \cite{ChenLuVishwanath2014,Devakuletal2018}.

Conjugating the first and second terms in the trivial Hamiltonian with the first and second expressions respectively, we obtain the Hamiltonian
\begin{align}
H = - \sum_{ij} &\left [  X(x^i y^j)Z \left (x^iy^j (1+ f y) s \right ) \right. \nonumber\\
  &\left. + X(x^i y^j s) Z \left (x^iy^j (1+ \bar f \bar y) \right )  \right].
\end{align}
Choosing the Sierpinski CA $f(x)= 1+x$ gives precisely the Hamiltonian for the cluster state on the honeycomb lattice \cite{Devakuletal2018,KubicaYoshida2018}.

\section{Pumping SSPTs protected by fractal symmetry}
\label{app:fractal}
We explicitly construct the unitaries that pump fractal SPTs to opposite sides of the bulk. Our 3D bulk is constructed by $L$ layers of 2D lattices, where each layer can be labeled by an additional index $z$ from $1$ to $z^L$. We define the gates
\begin{align}
V_{ijk} &= CZ \left ( x^iy^j z^k(1+z) , x^iy^j (1+ \bar f \bar y)z^k s \right),\\
V'_{ijk} &= CZ \left ( x^iy^j (1+fy)z^{k} , x^iy^j z^k(1+z)s    \right).
\end{align}
For $k=0,...,L-1$. These gates commute with the fractal symmetries extended weakly into the bulk
\begin{align}
S(q(x)) &=  X \left (q(x) \mathcal F(x,y) \sum_{l=0}^L z^l  \right),  \\
S'(q(x)) &=  X \left (q(x) \bar{\mathcal F}(x,y)  \sum_{l=0}^L z^l s \right ) .
\end{align}
That is, the symmetries in the bulk are products of identical fractal symmetries in all layers. Let us show this explicitly for $V_{ijk}$. The computation is identical for $V'_{ijk}$. 

\begin{table*}[t!]
     \centering

     \begin{tabular}{|c|c|c|c|}
          \hline
          Dim of cluster state & Bulk lattice & Topological stabilizer code & Transversal gate\\
     \hline
        1D  & Union-Jack / Triangular & Color code & $S$ \cite{KubicaBeverland15,Kesselringetal18}  \\
        1D  & Square & Two Toric codes & $CZ$ \cite{PotterMorimoto2017}\\
        2D  & FCC & Checkerboard & $S$ \cite{TantivasadakarnVijay2019}\\
        2D  & Cubic & X-cube & ??\\
        2D & Honeycomb stacks & Fractal spin liquid $\otimes$ Dual fractal spin liquid & $CZ$\cite{KubicaYoshida2018}\\
          \hline
     \end{tabular}
     \caption{Correspondence between cluster state pumps and transversal gates of the corresponding quantum codes after gauging the symmetries in the bulk.}
     \label{tab:transversalgate}
 \end{table*}

To show that $V_{ijk}$ and $S(q(x))$ commute, let $\alpha=q(x)\mathcal F(x,y) \sum_{l=0}^L z^l$, $\beta= (1+z)$ and $\gamma = x^iy^j (1+\bar f \bar y)z^k$. The commutation polynomial of $\alpha$ and $\beta$ is:
\begin{align}
P(\alpha,\beta) &=q(x)\mathcal F(x,y) \sum_{l=0}^L z^l (1+\bar z) \nonumber\\
&=q(x)\mathcal F(x,y) (\bar z + z^L).
\end{align}
Since $P_{ijk}=0$ for $k=0,...,L-1$, Eq. \eqref{equ:CZcomm1} implies that the two terms commute. To show that $V_{ijk}$ and $S'(q(x))$ commute, let $\alpha=q(x)\bar{\mathcal F}(x,y) \sum_{l=0}^L z^l$, $\beta= x^iy^j (1+ f y)z^k $ and $\gamma = (1+ \bar f \bar y)z^k$. The coefficients $P_{ijk}$ of the commutation polynomial
\begin{align}
P(\alpha,\gamma) &= q(x)\bar{\mathcal F}(x,y) \sum_l z^l (1+ f y) \bar z^k \nonumber \\
 &= q(x) \sum_l z^{l-k} . 
\end{align}
vanish for all $j>0$, so  Eq. \eqref{equ:CZcomm2} implies that the two terms commute.

We conclude that the unitary
\begin{align}
 U &= \prod_{ij}\prod_{k=0}^{L-1} V_{ijk} V'_{ijk}
\end{align}
acts trivially in the bulk, but creates cluster states at layers $z^0$ and $z^L$. This is can be achieved by evolving with the Hamiltonian
 \begin{equation}
H= \sum_{ij}\sum_{k=0}^{L-1} \left [V_{ijk} + V'_{ijk}\right]
 \end{equation}
 for time $\pi/2$.

 \section{Relation between pumps and transversal logical gates}
 
 There is a close connection between quantum pumps and transversal gates in topological codes. Recall that a transversal gate is a global action of an onsite operator which leaves the ground state subspace of the topological stabilizer code invariant. The ground state of this code can be related to that of a trivial product state by virtue of ``ungauging" \cite{CobaneraOrtizNussinov2011,VijayHaahFu2016,Williamson2016,Yoshida2017,KubicaYoshida2018,Pretko2018,ShirleySlagleChen2019,Radicevic2019,Tantivasadakarn20}. Under this mapping, the ``ungauged" transversal gate must also leave the trivial product state invariant, which means it acts as the identity.  Indeed, this is only true on a closed space manifold. When the manifold has a boundary, the ungauged operator can act non-trivially on the boundary, thus corresponding to a quantum pump. The converse also holds. Therefore, any quantum pump corresponds to a transversal gate in the gauged topological code. In Table \ref{tab:transversalgate}, we tabulate the correspondence between the various cluster state pumps presented, and the transversal gates in obtained after gauging. Note that all the transversal gates that prepare cluster states are Clifford. Indeed, one can show that transversal gates higher up in the Clifford hierarchy will ungauge to unitaries that pump hypergraph states to the boundary.
 
 We also remark that although transversal logical gates for toric codes and color codes have been extensively studied and classified \cite{KubicaBeverland15,Yoshida2015,Yoshida2017,Kesselringetal18,WebsterBartlett18}, such gates which would ungauge to 2D cluster state pumps are not well studied in fracton codes apart from a few examples\cite{KubicaYoshida2018,TantivasadakarnVijay2019}. Thus, a further study of these topics can simultaneously shed light on both measurement-based and topological quantum computation.

 \section{Bound for Perturbations in Driving Hamiltonian}
 \label{app:bound}
In the main text, we use feed-forward correction to guarantee the usability of the final resource state. In the absence of such correction, one can still use the final resource state upon post-selection. Here, we estimate the bound on the size of perturbations in the driving Hamiltonian needed to ensure a successful preparation of a cluster state on the boundary. For simplicity, we give an example for the setup on the square lattice, although it is straightforward to generalize to other 2D lattices and the preparation of 2D cluster states in 3D. In particular, we will show that the bound is  $\epsilon  <<N^{-1/2}$, where $N$ is the number of sites of the desired cluster state.

 We consider driving Hamiltonian in Eq. \eqref{equ:1DclusterHam} on a square lattice $M$ with boundary $\partial M$. Without perturbations, evolution using this Hamiltonian leads to the state
 \begin{equation}
 U\ket{\psi_0} = \ket{\text{cluster}}_{\partial M} \otimes \ket{+}_{M}
 \end{equation}
  To leading order, we can consider the following two types of symmetric perturbations: 
  \begin{align}
\Delta H^Z &= \epsilon\sum_{\Box_{1234}} Z_1Z_3 + Z_2Z_4,\\
\Delta H^X &= \epsilon\sum_i  X_i.
 \end{align}
Furthermore, we can consider the perturbations separately, since any mixed-terms would be higher order in $\epsilon$. We will demonstrate the calculation for $\Delta H^Z$, which is easier since it commutes with the Hamiltonian. The calculation for $\Delta H^X$ is nearly identical and will be briefly outlined.

Consider adding only $\Delta H^Z$. The resulting unitary to first order in $\epsilon$ is
  \begin{align}
U' &\approx U \prod_{\Box_{1234}} \left( 1+ i \frac{\pi}{2} \epsilon Z_1Z_3 \right)\left( 1+ i \frac{\pi}{2}  \epsilon Z_2Z_4 \right)\\
&\approx U \left[1+ i \frac{\pi}{2}\epsilon \sum_{\Box_{1234}} \left( Z_1Z_3+  Z_2Z_4 \right)   \right ]
 \end{align}
The final state to first order in order $\epsilon L$ is schematically
\begin{align}
 U'\ket{\psi_0} \approx& \left( \ket{\text{cluster}}_{\partial M} \otimes \ket{+}_{M} \right) \nonumber \\
 &+ i \frac{\pi}{2} \epsilon  \sum_{\Box \in \partial M}  \left( Z \ket{ \text{cluster}}_{\partial M} \otimes Z \ket{+}_{M} \right).
 \end{align}
Here, we have thrown away terms where the errors act only in the bulk, and did not affect the cluster state on the boundary since this does not affect our argument.

The final state is a superposition of the correct cluster state (tensored with a trivial bulk), with $\mathcal O(L)$ wrong states coming from each of the terms adjacent to the boundary. These wrong states have a single $Z$ operator on the cluster state (breaking the symmetry on the standalone boundary) and another $Z$ acting on the bulk, which changes one of the $\ket{+}$ states to $\ket{-}$. 

 We can collapse the wavefunction to the desired cluster-state by performing a projective measurement in the $X$-basis, and throw away all readouts with outcome $\ket{-}$. Since each wrong state can appear with probability of order $\epsilon^2$, the probability of failure is $\mathcal O (\epsilon^2 L)$. Thus, we can consistently create a 1D cluster state as long as $\epsilon << L^{-1/2}$.
 
It is worth pointing out that cluster-like states (for example $Z_i Z_j\ket{\text{cluster}}_{\partial M}$ where $i,j \in \partial M$) only appear at higher orders in $\epsilon$. Taking this into account, along with the fact that errors can also act purely in the bulk, we therefore only need to throw away measurements which result in an odd number of $\ket{-}$ states in the bulk. 

For the perturbation $\Delta H^X$, since the errors do not commute with our driving Hamiltonian, the unitary expanded to first order in $\epsilon$ is
\begin{equation}
U' = U e^{-i\epsilon\frac{\pi}{2} \sum_i X_i} e^{\frac{\pi^2}{8} \epsilon \sum_i [H, X_i]}.
\end{equation}
The errors do not come from the first exponential but from the commutation between $H$ and $X_i$ in the second exponential. Computing the commutator explicitly, one finds that the terms with support on both bulk and boundary will also flip a $\ket{+}$ to $\ket{-}$ in the bulk, and so can be eliminated by projective measurements and post-selection.

When generalizing to 2D, we see that the number of terms that can break the symmetry on the boundary is of order $L^2$, the number of sites on the boundary. Thus in general, the error only needs to be much smaller than $N^{-1/2}$ where $N$ is the number of sites on which we prepare the cluster state.

\bibliography{references}
\onecolumngrid
\clearpage
\twocolumngrid
\appendix